\documentclass[%
 reprint,
superscriptaddress,
 amsmath,amssymb,
 aps,
 prx,
]{revtex4-2}

\usepackage{chemformula} % Formula subscripts using \ch{}
\usepackage{xfrac}  % Works better with other fonts
\usepackage{float} 
\usepackage{physics}
\usepackage{bbold}
\usepackage{amsmath}
\usepackage{chemformula}
\usepackage{dsfont}
\usepackage{gensymb}
\usepackage{graphicx}% Include figure files
\usepackage{dcolumn}% Align table columns on decimal point
\usepackage{bm}% bold math
\usepackage{hyperref}% add hypertext capabilities
\begin{document}

\title{PyCCE: A Python Package for Cluster Correlation Expansion Simulations of Spin Qubit Dynamics}% Force line breaks with \\

\author{Mykyta Onizhuk}
\affiliation{Department of Chemistry, University of Chicago, Chicago, IL 60637, USA}
\affiliation{Pritzker School of Molecular Engineering, University of Chicago, Chicago, IL 60637, US}
\author{Giulia Galli}
\email[]{Corresponding author gagalli@uchicago.edu}
\affiliation{Department of Chemistry, University of Chicago, Chicago, IL 60637, USA}
\affiliation{Pritzker School of Molecular Engineering, University of Chicago, Chicago, IL 60637, US}
\affiliation{Materials Science Division and Center for Molecular Engineering, Argonne National Laboratory, Lemont, IL 60439, USA}

% \date{\today}% It is always \today, today,
%              %  but any date may be explicitly specified

\keywords{Spin Qubit, Coherence, Python}

\begin{abstract}
We present PyCCE, an open-source Python library to simulate the dynamics of spin qubits in a spin bath, using the cluster-correlation expansion (CCE) method. PyCCE includes modules to generate realistic spin baths, employing coupling parameters computed from first principles with electronic structure codes, and enables the user to run simulations with either the conventional or generalized CCE method. We illustrate three use cases of the Python library: the calculation of the Hahn-echo coherence time of the nitrogen vacancy in diamond; the calculation of the coherence time of the basal di-vacancy in silicon carbide at avoided crossings; and the magnetic field orientation-dependent dynamics of a  shallow donor in silicon.

The complete documentation and installation instructions are available at

\url{https://pycce.readthedocs.io/en/latest/}.

\end{abstract}

%\keywords{Suggested keywords}%Use showkeys class option if keyword
                              %display desired
\maketitle

%\tableofcontents

In solid-state systems, the spin of electrons is one of the leading platforms for quantum information applications \cite{Wolfowicz2021, Morello2020, PhysRevA.57.120}. 
Examples include but are not limited to quantum dot spin qubits \cite{Hendrickx2020, Bluhm2011, Watson2018, Petta2180}, shallow donors in Si \cite{Wolfowicz2013, Tyryshkin2012, Simmons2011}, and localized spin defects in semiconductors \cite{BarGill2013, Koehl2011, Bernien2013}.

All of these systems share one important attribute: the time evolution of the spin qubit, at least in part, is controlled by its magnetic environment, consisting of nuclear and other electronic spins. This environment can hinder the applications of spin-based technologies, imposing limits on the spin qubits' coherence times \cite{Yang_2016} but also be an advancement for some applications where nuclear spins may act as memory units in quantum networks \cite{pezzagna2021, Pompili259}. First principles predictions of the spin bath dynamics in materials are critical to the design of solid-state spin technologies and the realization of their full potential.

The cluster correlation expansion (CCE) method \cite{PhysRevB.78.085315, PhysRevB.79.115320} is one of the most widely used approaches  to simulate the quantum decoherence dynamics of spin qubits in a finite spin bath \cite{Yang_2016}. The approach reproduces the correct dynamics in a variety of physical realizations of spin qubits. For example, the results of the CCE method show excellent agreement with experiments for bismuth \cite{PhysRevB.89.045403, PhysRevB.91.245416} and phosphorous \cite{Ma2014} donors in Si, the nitrogen-vacancy (NV) center in diamond \cite{PhysRevB.85.115303}, and divacancies in 4H-SiC, both axial \cite{Seo2016, Bourassa2020} and basal \cite{PRXQuantum.2.010311}. The CCE method has also been used to predict the properties of novel materials, such as 2D platforms for spin qubits \cite{onizhuk_apl2021, Ye2019}, or conduct a general screening of possible qubit hosts over a wide range of materials \cite{kanai2021generalized}. However, the software implementation of the CCE method is not readily accessible and hinges on in-house developments, which lack openness and transferability.

To overcome these challenges, we developed \texttt{PyCCE} - an open-source python-based library for carrying out CCE calculations. This module is the first public implementation of the CCE code, available as an open-source package with high integrability within the existing scientific Python ecosystem. Here we describe  the theoretical framework of the CCE method and its implementation in the \texttt{PyCCE} library. We then highlight the application of the \texttt{PyCCE} module to several physical systems and verify and validate our implementation with theoretical and experimental data.

\section{Theoretical framework}
We begin by discussing the theoretical background of the model used in the CCE.

\subsection{Hamiltonian of system}
The Hamiltonian for the central spin in a  spin bath includes the following terms:

\begin{equation} \label{eq:totalh}
    \hat H = \hat H_S + \hat H_{SB} + \hat H_{B}
\end{equation}

Where $\hat H_S$ is the Hamiltonian of the free central spin, $\hat H_{SB}$ denotes interactions between central spin and a spin belonging to the bath, and $\hat H_B$ are intrinsic bath spin interactions: 

\begin{align}
    &\hat H_S = \mathbf{SDS} + \mathbf{B\gamma}_{S}\mathbf{S} \\
    &\hat H_{SB} = \sum_i \mathbf{S}\mathbf{A}_i\mathbf{I}_i \\
    &\hat H_{B} =  \sum_i{\mathbf{I}_i\mathbf{P}_i \mathbf{I}_i +
                   \mathbf{B}\mathbf{\gamma}_i\mathbf{I}_i} +
                   \sum_{i>j} \mathbf{I}_i\mathbf{J}_{ij}\mathbf{I}_j 
\end{align}

Where $\mathbf{S}=(\hat{S}_x, \hat{S}_y, \hat{S}_z)$ are the components of spin operators of the central spin, and $\mathbf{I}=(\hat{I}_x, \hat{I}_y, \hat{I}_z)$ are the components of the bath spin operators. The interactions are described by the following tensors:

\begin{itemize}
    \item $\mathbf{D}$ ($\mathbf{P}$) is the self-interaction tensor of the central spin (bath spin). For the electron spin, the tensor corresponds to the zero-field splitting (ZFS) tensor. For nuclear spins corresponds to the quadrupole interactions tensor. 
    \item $\mathbf{\gamma}_i$ is the magnetic field interaction tensor of the $i$-spin describing the interaction of the spin and the external magnetic field $\mathbf{B}$. We assume that for the bath spins, it is isotropic.
    \item $\mathbf{A}$ is the interaction tensor between central and bath spins. In the case of the nuclear spin bath, it corresponds to the hyperfine couplings.
    \item $\mathbf{J}$ is the interaction tensor between bath spins. 
\end{itemize}

\subsection{Qubit coherence}

Usually, two coherence times are measured to characterize the loss of a qubit coherence - $T_1$ and $T_2$. $T_1$ defines the timescale over which the qubit population is thermalized; $T_2$ describes a purely quantum phenomenon - the loss of the phase of the qubit's superposition state.

In the pure dephasing regime ($T_1 >> T_2$) the decoherence of the central spin is completely determined by the decay of the off diagonal element of the density matrix of the qubit. Namely, if the qubit is initially prepared in the $\ket{\psi}=\frac{1}{\sqrt{2}}(\ket{0}+e^{i\phi}\ket{1})$ state, the loss of the relative phase of the $\ket{0}$ and $\ket{1}$ levels is characterized by the coherence function:
\begin{equation}\label{eq:l_definition}
    \mathcal{L}(t) = \frac{\bra{1}\hat{\rho}_S(t)\ket{0}}{\bra{1}\hat{\rho}_S(0)\ket{0}}=\frac{\langle{\hat \sigma_-(t)}\rangle}{\langle{\hat \sigma_-(0)}\rangle}
\end{equation}
Where $\hat{\rho}_S(t)$ is the density matrix of the central spin, and $\ket{0}$ and $\ket{1}$ are qubit levels.

In general, one can potentially obtain the coherence function of the qubit by directly solving the Schrodinger equation of the total system and tracing out the bath degrees of freedom. However, the complexity of such computation grows exponentially with the bath size, and for any meaningful system (bath of several hundred spins), the problem is impossible to solve.

The core idea of the CCE approach is that the spin bath-induced decoherence can be factorized into set of irreducible contributions from the bath spin clusters. Written in terms of the coherence function:

\begin{equation}\label{eq:l_cce}
    \mathcal{L}(t) = \prod_{C} \Tilde{L}_C = \prod_{i}\Tilde{L}_{\{i\}}\prod_{i,j}\Tilde{L}_{\{ij\}}...
\end{equation}

Where $\Tilde{L}_{\{i\}}$ is the contribution of the single bath spin $i$, $\Tilde{L}_{\{ij\}}$ is an irreducible contribution of the spin pair $i,j$. The maximum size of the cluster included into the expansion determines the order of the CCE approximation. For example, in the CCE2 approximation, only contributions up to spin pairs are included, and in CCE3 - up to triplets of bath spins are included, etc. Each cluster contribution is defined recursively as:

\begin{equation}\label{eq:l_contribution}
    \Tilde{L}_C = \frac{L_{C}}{\prod_{C'}\Tilde{L}_{C'\subset C}}
\end{equation}

Where $L_{C}$ is a coherence function of the qubit, interacting only with the bath spins in a given cluster $C$, and $\Tilde{L}_{C'}$ are contributions of $C'$ subcluster of $C$.

The first element of the cluster expansion $\Tilde{L}_{\{0\}}=L_{\{0\}}$ is a phase factor of the free evolution of the central spin. Then the contribution of the single bath spin $i$ is computed as:

\begin{equation}
    \Tilde{L}_{\{i\}} = \frac{L_{\{i\}}}{\Tilde{L}_{\{0\}}}
\end{equation}

Next, the contribution of spin pairs is computed as:

\begin{equation}
    \Tilde{L}_{\{ij\}} = \frac{L_{\{ij\}}}{\Tilde{L}_{\{0\}}\Tilde{L}_{\{i\}}\Tilde{L}_{\{j\}}}
\end{equation}

And so on.

The workload of the iterative evaluation of correlation functions grows polynomially, instead of exponentially as a function of bath spins, making computations for thousands of bath spins trivial.

\begin{figure}
    \centering
    \includegraphics[scale=1]{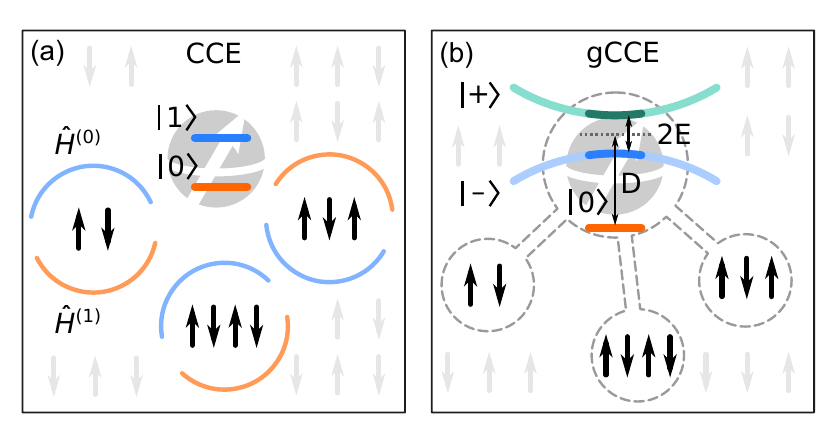}
    \caption{CCE Formulations. (a) Conventional CCE, where the cluster Hamiltonian is reduced to the sum of two effective Hamiltonians $\hat{H}^{(0)}$ and $\hat{H}^{(1)}$. (b) Generalized CCE, where each cluster Hamiltonian includes all central spin levels.}
    \label{fig:methods}
\end{figure}

\subsection{Conventional CCE}
In the original formulation of the CCE method (Fig. \ref{fig:methods}a), the total Hamiltonian of the system (Eq. \ref{eq:totalh}) is reduced to the sum of two effective Hamiltonians, conditioned on the qubit levels of the central spin:

\begin{equation}
    \hat H = \ket{0}\bra{0}\otimes\hat H^{(0)} + \ket{1}\bra{1}\otimes\hat H^{(1)}
\end{equation}

Where $\hat H^{(\alpha)}$ is an effective Hamiltonian acting on the bath when the central qubit is in the $\ket{\alpha}$ state ($\ket{\alpha}=\ket{0},\ket{1}$ is one of the two eigenstates of the $\hat H_S$ chosen as qubit levels):

\begin{equation}
    \hat H^{(\alpha)} = E_\alpha + \bra{\alpha}\hat H_{SB} \ket{\alpha} + \hat H_{B} + \hat H^{(\alpha)}_{PT}
\end{equation}

Where $\hat H^{(\alpha)}_{PT}$ are higher order perturbations to the effective Hamiltonian $\hat H^{(\alpha)}$. Up to the second order we can write:

\begin{equation}
    \hat H^{(\alpha)}_{PT} = \sum_{i,j} \sum_{\beta \neq \alpha} \frac{\left(\bra{\alpha}\mathbf{S}\ket{\beta}\mathbf{A}_i\mathbf{I}_i\right)     
           \left(\bra{\beta}\mathbf{S}\ket{\alpha}\mathbf{A}_j\mathbf{I}_j\right)} {E_\alpha - E_\beta} = \sum_{i,j} \mathbf{I}_i \mathbf{T_{ij}} \mathbf{I}_j
\end{equation}

Here summation goes over all $\ket{\beta}$ eigenstates of the central spin Hamiltonian $\hat H_S$.

For each cluster $C$ the Hamiltonian $\hat H_C^{(\alpha)}$ obtained by tracing out the degrees of freedom of the bath spins, not included in the cluster is:

\begin{equation}
\begin{split}\label{eq:conv_cce_ham}
    \hat H_C^{(\alpha)} = &\sum_{i\in C} \bra{\alpha} \mathbf{S} \ket{\alpha} \mathbf{A}_i \mathbf{I}_i + \sum_{i\in C} \mathbf{I}_i\mathbf{P}_i \mathbf{I}_i +\mathbf{B}\mathbf{\gamma}_i\mathbf{I}_i + \\ &\sum_{i,j \in C} \mathbf{I}_i ( \mathbf{J}_{ij} + \mathbf{T}_{ij}) \mathbf{I}_j + \sum_{i\in C,\ a\notin C} {\mathbf{I}_i( \mathbf{J}_{ia} + \mathbf{T}_{ia})\langle\mathbf{I}_a\rangle}
\end{split}
\end{equation}

Where $\langle\mathbf{I}_a\rangle=\Tr{\hat \rho_a\mathbf{I}_a}$ is an average value of each bath spin outside of a given cluster, and $\rho_a$ is the initial density matrix of the bath spin.

In most systems of interest, it is justified to assume that the state of each bath spin $i$ with spin $s$,is purely random ($\hat{\rho_i} = \frac{1}{2s+1}\mathds{1}$ , where $\mathds{1}$ is the identity). Then the $\langle\mathbf{I}_a\rangle=0$ in  Eq. (\ref{eq:conv_cce_ham}). However, this approximation is not always valid, and it may lead to incorrect results, as we discuss below.

Given an initial qubit state $\ket{\psi}=\frac{1}{\sqrt{2}}(\ket{0}+e^{i\phi}\ket{1})$ and an initial state of the bath spin cluster $C$ characterized by the density matrix $\hat \rho_{C}$, the coherence function of the qubit interacting with the cluster $C$ is computed as:

\begin{equation}
    L_{C}(t) = Tr[\hat U_C^{(0)}(t)\hat \rho_C \hat U_C^{(1) \dagger}(t)]
\end{equation}

Where $\hat U_C^{(\alpha)}(t)$ is time propagator defined in terms of the effective Hamiltonian $\hat H_C^{(\alpha)}$ and the number of decoupling pulses. For free induction decay (FID) the time propagators are trivial:

\begin{equation}
    \hat U_C^{(0)} = e^{-\frac{i}{\hbar} \hat H_C^{(0)} t};\
    \hat U_C^{(1)} = e^{-\frac{i}{\hbar} \hat H_C^{(1)} t}
\end{equation}

Each applied $\pi$-pulse flips the state of the central spin, and therefore changes the evolution of the bath. For the Hahn-echo sequence (where a single $\pi$-pulse is applied halfway between the initialization and the measurement of the qubit) we can write the propagators as:
\begin{align}
    \hat U^{(0)} &= e^{-\frac{i}{\hbar} \hat H_C^{(1)} \frac{t}{2}}
                    e^{-\frac{i}{\hbar} \hat H_C^{(0)} \frac{t}{2}} \\
    \hat U^{(1)} &= e^{-\frac{i}{\hbar} \hat H_C^{(0)} \frac{t}{2}}
                    e^{-\frac{i}{\hbar} \hat H_C^{(1)} \frac{t}{2}}
\end{align}

And for the generic decoupling sequence with $N$ (even) decoupling pulses applied at $t_1, t_2...t_N$, we write:
\begin{equation}
    \hat U^{(\alpha)}(t) = e^{-\frac{i}{\hbar} \hat H_C^{(\alpha)} \Delta t_{N}}
                           e^{-\frac{i}{\hbar} \hat H_C^{(\beta)} \Delta t_{N-1}}
                           ...
                           e^{-\frac{i}{\hbar} \hat H_C^{(\alpha)} \Delta t_{1}}
\end{equation}
Where $\ket{\alpha} = \ket{0}, \ket{1}$ and $\ket{\beta} = \ket{1}, \ket{0}$ accordingly. $\Delta t_{n} = t_{n} - t_{n-1}$ is the time difference between consequent pulses,
$t=\sum_i{t_i}$ is the total evolution time. In sequences with odd number of pulses $N$, the leftmost propagator is the exponent of $\hat H_C^{(\beta)}$.
For further details, we refer the readers to the seminal papers on CCE \cite{PhysRevB.78.085315, PhysRevB.79.115320}.

\begin{figure*}
    \centering
    \includegraphics[scale=1]{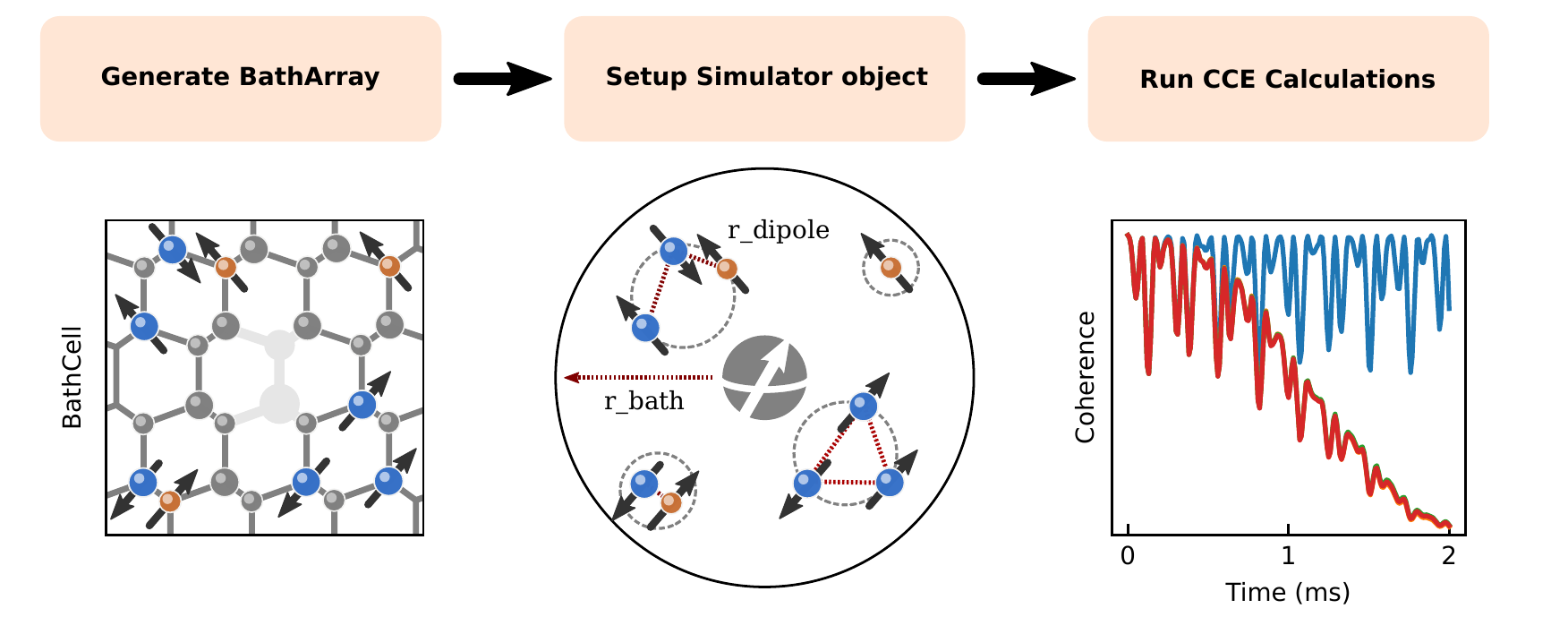}
    \caption{General workflow of the \texttt{PyCCE} module. From left to right: generate the spin bath; determine the properties of the central and cluster composition; run CCE simulations.}
    \label{fig:workflow}
\end{figure*}

\subsection{Generalized CCE (gCCE)}

Instead of projecting the total Hamiltonian on the qubit levels, one may directly include the central spin degrees of freedom to each clusters. We refer to such formulation as gCCE (Fig. \ref{fig:methods}b). In this case we write the cluster Hamiltonian as:

\begin{equation}\label{eq:gcce_H_C}
\begin{split}
    \hat H_C & {} =  \mathbf{SDS} + \mathbf{B\gamma_{S}}\mathbf{S} + \sum_{i\in C} \mathbf{S} \mathbf{A}_i \mathbf{I}_i + \sum_{i\in C} \mathbf{I}_i\mathbf{P}_i \mathbf{I}_i +\mathbf{B}\mathbf{\gamma}_i\mathbf{I}_i +  \\ & \sum_{i<j \in C} \mathbf{I}_i \mathbf{J}_{ij} \mathbf{I}_j +\sum_{a \notin C} \mathbf{S} \mathbf{A}_a \langle\mathbf{I}_a\rangle + \sum_{i\in C,\ a\notin C} {\mathbf{I}_i\mathbf{J}_{ia}\langle\mathbf{I}_a\rangle}
\end{split}
\end{equation}

And the coherence function of the cluster $L_C(t)$ is computed as:

\begin{equation}\label{eq:gcce_l}
    L_{C}(t) = \bra{0}\hat U_C(t)\hat \rho_{C+S} \hat U_C^{\dagger}(t)\ket{1}
\end{equation}

Where $\hat \rho_{C+S} = \hat \rho_{C} \otimes \hat \rho_S $ is the combined initial density matrix of the bath spins' cluster and central spin.

Similar to thee conventional CCE, we define the time propagator in terms of number of decoupling $\pi$-pulses and the cluster Hamiltonian (Eq. \ref{eq:gcce_H_C}). FID propagator is given as:

\begin{equation} \label{fid_propagator}
\hat U_C(t) = e^{-\frac{i}{\hbar}\hat H_C t}
\end{equation}

For an arbitrary set of decoupling pulses we can write the propagator as follows:

\begin{equation} \label{dd_propagator}
\hat U_C(t) = 
\mathcal{T} \left[e^{-\frac{i}{\hbar} \hat H_C \tau } e^{-\frac{i}{\hbar} \hat\sigma_{\{x, y, z\}} \frac{\phi}{2}} e^{-\frac{i}{\hbar} \hat H_C  \tau}\right]^N
\end{equation}

where $\hat\sigma_{\{x,y,z\}}$ are the Pauli matrices of the qubit, $\tau$ is the delay between pulses, $\phi$ is the rotational angle, and $N$ is number of pulses. For Hahn-echo experiments with a $\pi$ rotation about the $x$ axis the propagator is given by:

\begin{equation} \label{he_propagator}
\hat U_C^{HE}(t) = e^{-\frac{i}{\hbar}\hat H_C\frac{t}{2}}e^{-\frac{i}{\hbar}\hat\sigma_{x} \frac{\pi}{2}} e^{-\frac{i}{\hbar} \hat H_C \frac{t}{2}}
\end{equation}

For further details, we refer the reader to ref. \cite{PRXQuantum.2.010311}.

\subsection{Monte Carlo bath state sampling}

To improve the convergence of the cluster expansion for some particularly challenging systems, we can directly sample the pure states of the bath to predict the random state of the bath. Then, the coherence is computed as:

\begin{equation}
    \mathcal{L}(t) = \sum_i p_i\mathcal{L}_i(t)
\end{equation}

Where $\mathcal{L}_i(t)$ is the coherence function, computed for the pure bath state $i$, and $p_i$ is the probability of such state (for the completely mixed bath state, all $p_i$ are equal).

\subsection{Correlation Function}

A cluster expansion similar to the one just described may be applied to the autocorrelation of the Overhauser field upon the central spin. The autocorrelation in secular approximation is given by:

\begin{equation}\label{eq:corr}
    \mathcal{C}_{AA}(t) = \left\langle\sum_{\{I\}} A_{zz}\hat I_z (t)\sum_{\{I\}} A_{zz}\hat I_z (0)\right\rangle
\end{equation}

Where the $\hat I_z (t)$ is the spin operator in the Heisenberg picture $\hat I_z (t)=\hat U^\dagger(t) \hat I_z \hat U(t)$.
Under the CCE approximation, we can write the autocorrelation function as a sum of the spin cluster contributions:

\begin{equation}
    \mathcal{C}_{AA}(t) = \sum_{C} \Tilde{C}_{AA,\ C}
\end{equation}

Where cluster contributions are defined recursively, in a fashion similar to coherence function contributions:

\begin{equation}
    \Tilde{C}_{AA,\ C} = {C}_{AA,\ C} -  \sum_{C'} \Tilde{C}_{AA,\ C' \subset C}
\end{equation}

And the noise autocorrelation function for each cluster is computed using  Eq. \ref{eq:corr} for the bath spins within the cluster. Further details are available in ref. \cite{PhysRevB.90.115431, PhysRevB.92.161403}.

\begin{figure}
    \centering
    \includegraphics[scale=1]{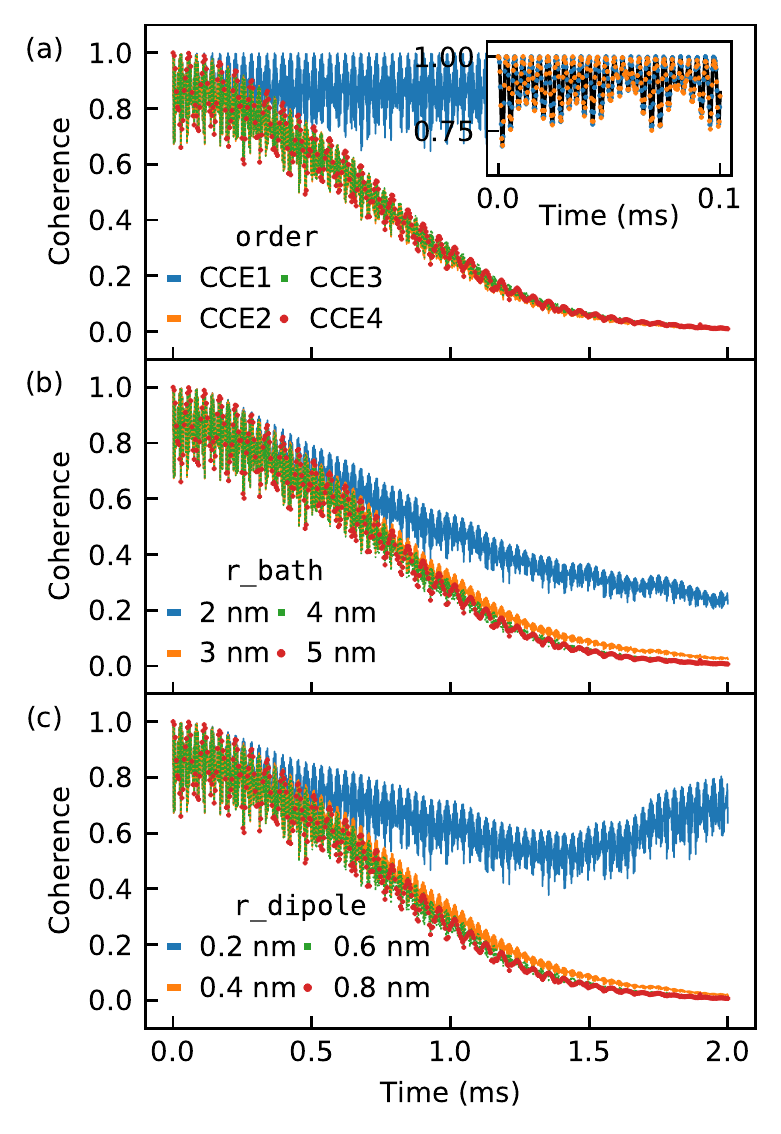}
    \caption{Convergence of the Hahn-echo coherence curve for \ch{NV^-} Center in diamond. (a) Convergence with respect to the CCE order. Inset shows CCE1 (blue) and CCE2 (orange) curves against the exact solution for the isolated nuclear spins. (b) Convergence of the size of the bath in angstrom. (c) Convergence of the connectivity distance in angstrom.}
    \label{fig:nv}
\end{figure}

\section{Code implementation}
In the \texttt{PyCCE} library, we provide a convenient way to simulate the dynamics of the central spin interacting with a realistic spin bath.

The general workflow of running simulations with \texttt{PyCCE} includes the following steps (Fig. \ref{fig:workflow}):

\begin{itemize}
\item Generate the spin bath as an instance of \texttt{BathArray} class.
\item Input the properties of the central spin to the instance of the \texttt{Simulator} class - central object used for CCE simulations.
\item Calculate the desired properties using the \texttt{Simulator} object.
\end{itemize}

\textbf{Generation of the spin bath}. The \texttt{BathCell} class allows one to initialize the structure of the material, populate it with spins and generate a large supercell, which is then stored in the \texttt{BathArray} format. The same object stores the properties of the bath spins (hyperfine couplings, quadrupole interactions, gyromagnetic ratios). Most of the properties of the nuclear spins (e.g., gyromagnetic ratio, concentrations) are already available in the \texttt{PyCCE} library using the EasySpin \cite{stoll2006} database.

The \texttt{BathCell} supports an interface with the Atomic Simulations Environment (ASE) package \cite{ase_2017}, a well-developed tool to initialize the structure of many solid-state materials.

\textbf{Setting up the \texttt{Simulator} object}. 
In this step, a user should provide properties of the central spin, including zero-field splitting, magnetic field interaction tensor, and the amplitude of the external magnetic field. 

Next, one chooses the size of the "active" bath spins by imposing a cutoff radius, \verb`r_bath`, defined as the maximum distance from the central spin to the bath spin to be considered in the calculation. 

The hyperfine couplings entering Eq. (\ref{eq:totalh}) are either: 

\begin{itemize}

\item{Approximated using the point dipole-dipole approximation as:

\begin{equation}
    \mathbf{A} = -\mathbf{\gamma_S} \gamma_i \frac{\mu_0 \hbar^2}{4\pi|\vec{r}|^5}
                 \left[3 \cdot \vec{r}\otimes\vec{r}  - \mathbb{1}\cdot|\vec{r}|^2 \right]
\end{equation}
Where $\mathbb{1}$ is $3\cross3$ identity matrix and the $\gamma_i$ tensors are assumed to be isotropic.
}
\item{Computed using the spin density distribution of the central spin. The \texttt{PyCCE} package supports input of the spin density of the central spin in the Gaussian cube format \cite{cube} and can integrate it to obtain dipolar hyperfine couplings.}

\item{Obtained from first principles calculations using  quantum chemistry or solid state simulation packages. As of now, an interface with ORCA \cite{neese2012} and Quantum Espresso \cite{Giannozzi2009} packages is provided in the \texttt{PyCCE} implementation.}

\item{Manually set by user}.

\end{itemize}
The hyperfine couplings are required for each bath spin within the "active" region. 

The interaction tensors between bath spins $\mathbf{J}_ij$ are assumed from point dipole-dipole interaction or can be provided by the user.

Finally, the clusters entering the cluster expansion are determined.
We followed the procedure of Ref. \cite{Seo2016, Yang_2016} to obtain the clusters entering Eq. (\ref{eq:l_cce}). We define a cutoff radius, \verb`r_dipole`, that sets the maximum distance at which two bath spins form an "edge." Bath spins $i$ and $j$ form a cluster of two if there is an edge between them (distance $d_{ij} <\verb`r_dipole`$). Bath spins $i$, $j$, and $k$ form a cluster of three if enough edges are found that connect them (e.g., there are two edges $ij$ and $jk$) and so on. In general, we assume that spins $\{i..n\}$ form clusters if they form a connected graph. Only clusters up to the size imposed by the \ verb'order` parameter (equal to the CCE order) are included.

\textbf{Running CCE simulations}.
Once all parameters of the Hamiltonian are set, the \verb`Simulator` object can be used to compute the coherence function of the central spin and thee autocorrelation function of the bath spin noise using both the conventional CCE and gCCE.

Full documentation is available online at \cite{documentation}.

\section{Showcasing capabilities}

\begin{figure}
    \centering
    \includegraphics[scale=1]{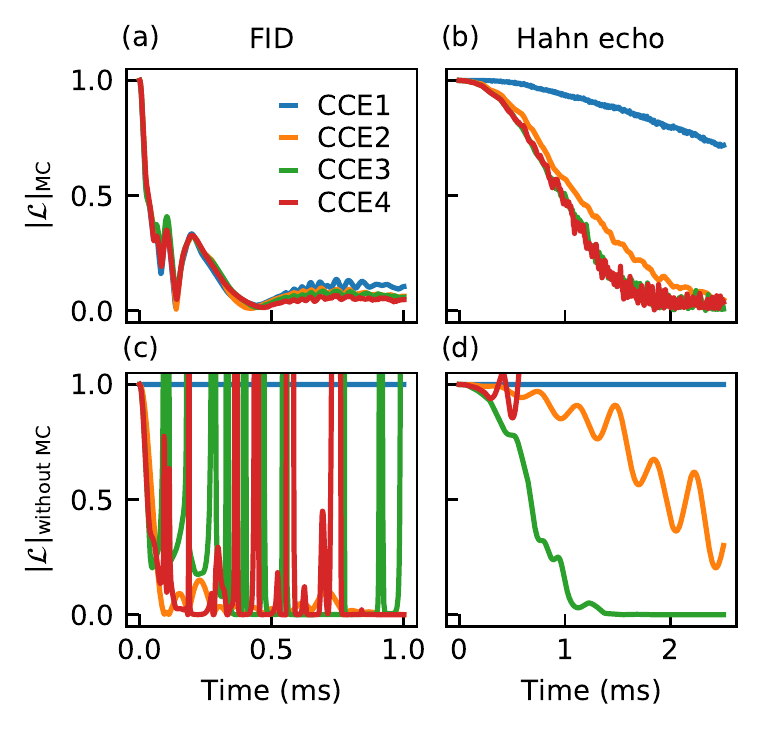}
    \caption{Coherence function of the basal kh-divacancy in 4H-SiC at avoided crossing of energy levels, computed with ($\mathcal{L}_\text{MC}$) or without ($\mathcal{L}_\text{without\ MC}$) bath state sampling for free induction decay ((a), (c)) and hahn echo ((b), (d)) experiments at different CCE orders.}
    \label{fig:basal}
\end{figure}

\subsection{NV Center in diamond}
First, we use \texttt{PyCCE} to highlight the application of the CCE method to a well-known physical system, the negatively charged nitrogen-vacancy (\ch{NV^-}) center in diamond. For this example, we compute hyperfine couplings assuming interactions between magnetic point dipoles.
In Figure \ref{fig:nv} we present the convergence of the ensemble-averaged decay of the Hahn-echo coherence function, computed with the conventional CCE.

The coherence function has two distinct elements - the fast electron spin echo envelope modulations (ESEEM) and the slow decay of the envelope. The calculations at the CCE1 level already reproduce the ESEEMs for short timescale, and exactly follow the analytical solutions for the bath of the isolated nuclear spins \cite{schweiger2001principles, Seo2016} (inset of Fig. \ref{fig:nv}a, black):
\begin{equation}
    \mathcal{L}^\text{exact}(t) = \prod_i 1 - 2 k_i 
        \sin^2{(w_i^{(0)}\frac{t}{4})} \sin^2{(w_i^{(+1)}\frac{t}{4})}
\end{equation}

Where $w_i^{(0)}=w_L$, $w^{(1)}=\sqrt{(w_L+A_{||})^2+A_{\bot}^2}$ are conditioned Larmor frequencies and
$k_i=\frac{A_{\bot}^2}{(w_L+A_{||})^2+A_{\bot}^2}$. Perpendicular and parallel hyperfine couplings are given as $A_{\bot}^2=A_{zx}^2+A_{zy}^2$, $A_{||}=A_{zz}$ respectively, and the free Larmom frequency is $w_L=\gamma_i B_z$ for $i$ nuclear spin.
The decay of the coherence function can be seen at the CCE2 level and higher orders.

We find that the coherence time $T_2$, obtained from the fit of the coherence function envelope to the compressed exponent $\exp[-(t/T_2)^n]$, is about 0.92 ms, in line with previously reported theoretical predictions \cite{PhysRevB.90.075201, PhysRevB.85.115303}. Finally, we confirm that CCE calculations converge at second order, size of the bath of $\approx4$ nm, and connectivity distance in the clusters of $\approx 0.6$ nm, consistent with supplementary information of ref. \cite{Seo2016}.

\subsection{Basal VV in SiC}
Next, we highlight the importance of the bath state sampling to converge the dynamics of the localized spin defects at avoided crossings. As an example, we consider the basal kh-divacancy in 4H-SiC. In this system, both nonzero axial $D$ and transverse $E$ zero-field splitting are non zero, leading to an anisotropic $\mathbf{D}$ tensor in Eq. (\ref{eq:totalh}). The presence of the anisotropy in the $\mathbf{D}$ tensor allows for avoided crossing to emerge at zero applied magnetic field. When operating the qubit at this avoided crossing, one may significantly prolong the spin coherence.

We consider one random nuclear spin configuration and compute the coherence function corresponding to the free induction decay (FID), and Hahn echo experiments, using thee gCCE method with and without Monte Carlo state sampling (Fig. \ref{fig:basal}a-d). We use \textit{ab initio} hyperfine couplings for the inner shell (at  distances $\leq$ 1 nm from the defect), computed with the GIPAW \cite{GIPAW} module of the Quantum Espresso package \cite{Giannozzi2009}. We used a plane-wave basis with a kinetic energy cutoff of 40 Ry and the Perdew-Burke-Ernzerhof (PBE) in DFT calculations. For the outer shell (distances $\geq$ 1 nm), we employed a point dipole approximation.

We find that the bath state sampling is crucial for the convergence of the coherence function for both free induction and Hahn-echo experiments. The coherence function is well converged at the second order in both cases. The decay of the coherence function is greatly dependent on the specific nuclear spin configuration, as discussed in Ref.\cite{PRXQuantum.2.010311}.

\begin{figure}
    \centering
    \includegraphics[scale=1]{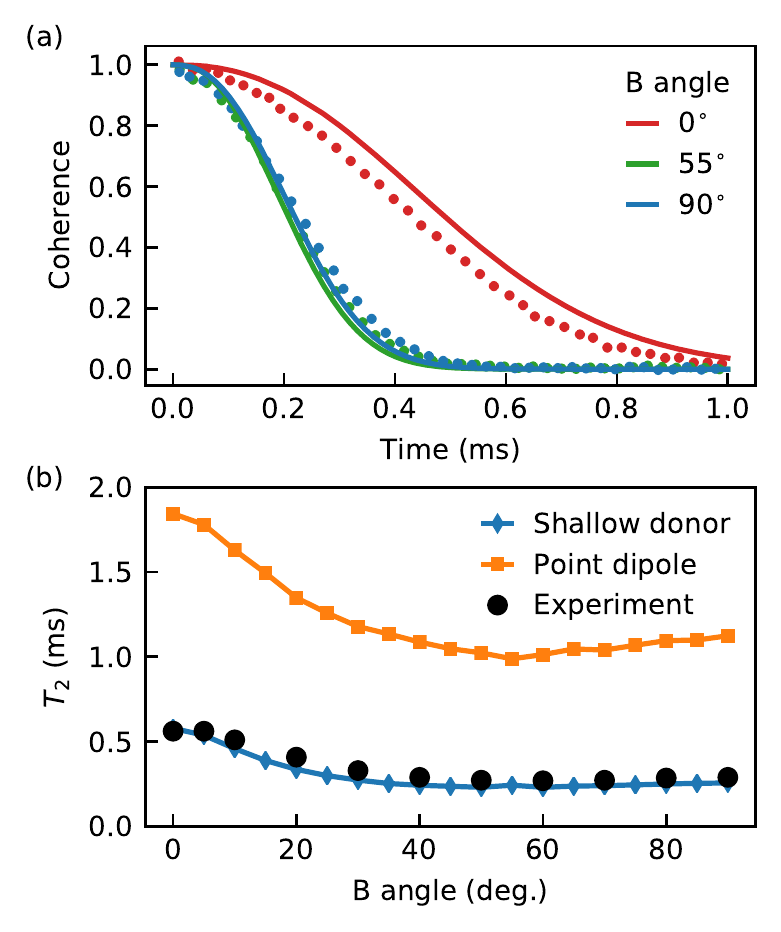}
    \caption{Coherence of the P donor in Si as a function of the magnetic field orientation, where $0\degree \rightarrow B \parallel [001]$, $55\degree \rightarrow B \parallel [111]$ and $90\degree \rightarrow B \parallel [110]$. (a) The coherence at three representative angles. Solid lines show the coherence, computed with \texttt{PyCCE}, and points show experimental data. (b) Computed coherence time $T_2$ as a function of the magnetic field rotation about $[1\Bar{1}1]$ axis for P shallow donor (blue) and for the hypothetical localized electron spin with the same nuclear spin bath (orange) against experimental results (black). Experimental data is taken from \cite{PhysRevB.82.121201}.}
    \label{fig:shallow}
\end{figure}

\subsection{Shallow P donor in Si}
Finally, we use \texttt{PyCCE} to compute the coherence decay of a shallow donor in Si induced by the nuclear spin bath. In this system, the contact Fermi terms dominate the hyperfine interactions.
The contact hyperfine couplings for \ch{^{29}Si} nuclear spins can be approximated using the Kohn-Luttinger wave function as:

\begin{equation}
\begin{split}
    A_F = \frac{16 \pi}{9} \gamma_S \gamma_{^{29}Si}\eta [&F_1(\vec{r})\cos{k_0 x} + F_3(\vec{r})\cos{k_0 y} + \\ &F_5(\vec{r})\cos{k_0 z}]^2
\end{split}
\end{equation}

Where the envelope functions $F_j$ are given as:

\begin{equation}
    F_1 = \frac{\exp{\left[ \sqrt{\frac{x^2}{(nb)^2}+\frac{y^2+z^2}{(na)^2}}\right]}}{\sqrt{\pi(na)^2(nb)}}
\end{equation}

Here $\vec{r}=x\vec{i}+y\vec{j}+z\vec{k}$ is the position of the nuclear spin, $k_0=0.85 \frac{2\pi}{a_{Si}}$, $a_{Si}=0.543$ nm is the lattice parameter, $n=0.81$ for a phosphorus impurity, $\eta = 186$ is the charge density at Si lattice site, $a=2.509$ nm, $b=1.443$ nm are characteristic lengths for hydrogenic impurities in Si \cite{PhysRevB.68.115322}.

The dipolar interaction is assumed to be equal to the one of point dipole at distances $\ge na\approx 2$ nm and zero otherwise \cite{Ma2014}.

The computed coherence for different alignments of the applied magnetic field is shown in Figure \ref{fig:shallow} along with the experimental results of Eisuke Abe et al. \cite{PhysRevB.82.121201}.

We find excellent agreement with the reported experiments as a function of the angle of the magnetic field. It is interesting to note that for the localized electron spin, such as a T-center \cite{PRXQuantum.1.020301}, we predict the coherence time in naturally abundant Si to be more than twice as that of the shallow donor (Fig. \ref{fig:shallow}b).

\section{Conclusion}
In this work, we presented an open-source python library to perform CCE calculations of the central spin interacting with a multitude of bath spins, including   
localized spin defects and shallow donors, as well as novel emerging platforms, such as molecular spin qubits \cite{Shiddiq2016, Bayliss1309}.

We verified the implementation of the method using theoretical data from previous studies and we validated several of our results with experiments.
In the future, we will extend the framework to include an arbitrary number of central spins. Implementation of so-called "hybrid CCE" \cite{Yang_2016, PhysRevB.86.035452} formulation, which allows one to study systems with the same type of bath spins as central spin, is also underway.

The open-source nature of our module allows for broad collaborations within the scientific community and will facilitate rapid advances in first principle predictions of the coherence dynamics of the spin qubits.

\section{Acknowledgements}

This work was supported by MICCoM, as part of the Computational Materials Sciences Program funded by the U.S. Department of Energy. This work made use of resources provided by the University of Chicago’s Research Computing Center.
We thank He Ma, Wennie Wang, Yu Jin, Nan Sheng, Maya Watts, and 
Yuxin Wang for helpful feedback on the code.

\bibliography{references}% Produces the bibliography via BibTeX.

\end{document}